\documentclass[12pt,a4paper]{article}
\usepackage{latexsym}
\usepackage{amsfonts}
\usepackage{multicol}
\usepackage[dvips]{graphicx}
\usepackage{fixltx2e}

\newcommand{\danger}[1]{\textbf{#1}}

\input{epsf}             % encapsulated Postscript figures
\begin{document}

\title{\danger{Black Hole Entropy in Loop Quantum Gravity and Number Theory}}
\author{\centerline{\danger{J. Manuel Garc\'\i a-Islas \footnote{
e-mail: jmgislas@leibniz.iimas.unam.mx}}}  \\
Instituto de Investigaciones en Matem\'aticas Aplicadas y en Sistemas \\ 
Universidad Nacional Aut\'onoma de M\'exico, UNAM \\
A. Postal 20-726, 01000, M\'exico DF, M\'exico\\}

\maketitle

\begin{abstract}
We show that counting different configurations that give rise to black hole entropy in loop quantum gravity is related to partitions in number theory. 
\end{abstract}

\section{Introduction}

The microscopic description of the Bekenstein-Hawking entropy \cite{b}, \cite{h} 
is one of the most important problems any theory of quantum gravity should explain.

In loop quantum gravity black hole entropy has been studied well for isolated horizons and 
of large area. One of the most fundamental problems for completing the task is to 
know exactly how many different configurations we have that give rise to a fixed 
area. 

There is an effective method for counting configurations by using methods of number theory
which was studied in \cite{abpbv}.  

In this paper we show that there is in fact a deeper relation between the counting of configurations
which give rise to black hole entropy and certain counting in number theory. 

We start our solution by considering first the simplest case which is given when the
area spectrum is equidistant. This situation for example emerges naturally in loop quantum gravity when we are
considering
very large spins.

In loop quantum gravity spin network states are eigenvalues of the area operator. The spin network
edges are labelled by half-integers $\{ j \in 0, 1/2, 1, ...\}$. When a surface is punctured
by an edge labelled with a spin $j$
the surface acquires the area  $A_j =8 \pi \gamma {\l_{p}}^2 \sqrt{j (j+1)}$,
where $\l_{p}$ is the Planck length and $\gamma$ is a parameter known by the name of Immirzi. 

\bigskip

More specifically, label the edges of
the spin network by $j_{i} $ which are half integers, that is, irreducible representations
of the group $SU(2)$.
Suppose the spin network punctures the surface in $n$ isolated points, and in a non-degenerate way.
Consider units for which $4 \pi \gamma {\l_{p}}^2=1$.
The total area of the surface is given by the eigenvalues of the area operator

\begin{equation}
A=  2 \sum_{i=1}^{n} \sqrt{j_{i}(j_{i}+1)}
\end{equation}

Now, if we have an isolated black hole, the microscopic description of its entropy is given by states
which live in the horizon surface. This entropy is
given by the logarithm of the number $\mathcal{N}$ of microstates which account for a fixed area of the surface.

The number of microstates is given as follows.
It is known that each spin network edge labelled $j$ which punctures the horizon contributes to the dimension of the boundary Hilbert space of states by a factor $(2j+1)$ which is the dimension of 
the irreducible representation $j$ of $SU(2)$. When considering all the punctures of the horizon the
dimension of the Hilbert space is given by the product of all the numbers $(2j+1)$ associated to
the spins which label the puncturing edges. 

The entropy is given by the logarithm of the dimension of the Hilbert space of the boundary.

\begin{equation}
S= \ln \ \mathcal{N}
\end{equation}
The problem we consider here is the counting of configurations which account for a fixed area
of the black hole horizon.
We also should take into account whether we are considering counting distinguishable
or indistinguishable configurations. 

A configuration is a set of edges of a spin network puncturing the horizon in a non-degenerate way 
and labelled $\{ n_{j} \} =\{ n_{1/2}, n_{1},...,n_{s_{max}/2} \}$ where $n_{j}$ is the number
of punctures with spin $j$, and where the following equation is satisfied 

\begin{equation}
A=  2 [n_{1/2} \sqrt{j_{1/2}(j_{1/2}+1)} +  n_{1} \sqrt{j_{1}(j_{1}+1)}
+...+ n_{s_{max}/2} \sqrt{j_{s_{max}/2}(j_{s_{max}/2}+1)}]
\end{equation}
We can ask equation 
$(3)$ to be satisfied exactly or we can also ask for configurations which area eigenvalue lies in an interval $[A-\delta, A+\delta ]$.  How many configurations satisfy equation $(3)$?

The problem of counting the number of these configurations started interestingly with ideas
of \cite{cr},  \cite{kk1}. Then it also
has been considered for example in 
 \cite{abpbv},
 \cite{abck}, \cite{dl}, \cite{cpb}, \cite{t}, \cite{tt}; however the states which account for
the entropy vary in opinions and we have various possibilities.    

For example, it has also been discussed whether two states which may vary by a permutation
of the same set of spins labeling edges of a fixed spin network should be considered
equivalent or not, see for example \cite{kk1}. The counting can be done for both possibilities 
as explained in \cite{kk1}. 

Here we consider the situation in which the counting is only related to different
(indistinguishable) 
configurations.  Two configurations which vary by a permutation of spins on the edges of a fixed
spin network are considered to be equivalent and are counted only once.
We do not worry about other quantum numbers which may be assigned
to the punctures such as half integers $m_{I}$ , such that $-j \leq m_{I} \leq j$ with the projection
$\sum_{I} m_{I}=0$.

\section{The counting}

The horizon has very large area. First consider the case 
of equidistant area spectrum, that is 
$A=  2 \sum_{i=1}^{n} (j_{i}+1/2)$.\footnote{Recall we are using units $4 \pi \gamma {\l_{p}}^2=1$}
This means that the spectrum becomes
equally spaced. In fact the case of equidistant spectrum as a serious candidate for the real spectrum
of loop quantum gravity 
has been considered
for example in \cite{kk}  \cite{aps}, \cite{p}, \cite{gs}.

In \cite{c} some arguments against the equally spaced spectrum are given due to inconsistency.

But recently it has been shown that we have to consider the equally spaced spectrum
as a serious issue \cite{blv} as flux-area operator with this property are shown to exist.
In \cite{s} the entropy of a black hole is also studied in terms of the equally spaced
spectrum. However as the lowest spin which contributes to the area 
in the case of $\sqrt{j(j+1)}$ is $1/2$ we stick to this situation. 

In this case equation $(3)$ can be stated as

\begin{equation}
A=  2 [n_{1/2} (j_{1/2}+1/2) +  n_{1} (j_{1}+1/2)
+...+ n_{s_{max}/2}(j_{s_{max}/2}+1)]
\end{equation}
which is equivalent to

\begin{equation}
A=  n_{1/2} (2j_{1/2}+1) +  n_{1} (2j_{1}+1)
+...+ n_{s_{max}}(2j_{s_{max}/2}+1)
\end{equation}
where all the numbers $n_{j}$ and $(2j +1)$ are integers.  Let $(2j+1)=m_j$. For example 
$m_{1/2}= 2$ where we assume $j=1/2$ to be the lowest value a spin can have. 
How many configurations satisfy
equation $(5)$ exactly? The question translates in calculating how many configurations 
$\{n_j\}$ where $m_{j}=(2j+1)$ belonging to the natural numbers exist such that

\begin{equation}
A=  n_{1/2} m_{1/2} +  n_{1} m_{1}
+...+ n_{s_{max}} m_{s_{max}/2}
\end{equation}
where $A$ is a natural number.

Now we describe how to do the exact counting of different and indistinguishable configurations which satisfy
equation $(6)$.  

\bigskip

\danger{Definition} A partition of a number $N$ is $N=\lambda_1 + \lambda_2 +...+\lambda_k $,
where $\lambda_1 \geq \lambda_2 \geq.....\geq \lambda_k \geq 1$. The summands are called the
parts of $N$.  The number of different partitions of $N$ is denoted $p(N)$. \cite{a}

\bigskip

Let us check that any configuration $\{ n_{j} \} =\{ n_{1/2}, n_{1},...,n_{s_{max}/2} \}$ which satisfies
equation $(6)$ is a partition of the number $A$ according to our definition. 
Recall that a configuration $\{ n_{j} \} =\{ n_{1/2}, n_{1},...,n_{s_{max}/2} \}$ of our isolated horizon 
is a sequence of numbers, where each $n_{j}$ only refers to the number of punctures with spin $j$.
The numbers $m_{j}=2j+1$ are really of importance since they refer to the dimension of the irreducible representations of $SU(2)$ 
of the spin $j$. 

We therefore think of formula $(6)$ as a partition of the number $A$ in which its parts are given
by the numbers
$m_{j}$.

Just note that formula $(6)$ is a sum given by

\bigskip

$A=  (m_{1/2} +  m_{1/2} +...+ m_{1/2}) +  (m_{1} + m_{1} +...+m_{1}) +......$

$....+ (m_{s_{max}/2} + m_{s_{max}/2} +...+m_{s_{max}/2})$

\bigskip

where the first bracketed sum of $m_{1/2}$ contains $n_{1/2}$ terms, 
the second bracketed sum of $m_{1}$ contains $n_{1}$ terms, and so on, such
that the last bracketed sum of $m_{s_{max}/2}$ contains $n_{s_{max}/2}$ terms.
According to the definition of a partition, the integer area $A$ needs to be decomposed as
$A= \lambda_1 + \lambda_2 +...+\lambda_k$, with 
$\lambda_1 \geq \lambda_2 \geq.....\geq \lambda_k \geq 1$;
it is now easy to see that we can take $\lambda_1= m_{s_{max}/2}$, 
$\lambda_2= m_{s_{max}/2}$ ,..., $\lambda_{n_{s_{max}/2}}= m_{s_{max}/2}$, where 
$n_{s_{max}/2} << k$.
We continue in this way till we get to
$\lambda_{k-n_{1/2}}= m_{1/2}$ ,...,  $\lambda_{k-1}= m_{1/2}$ , $\lambda_{k}= m_{1/2}$. 
\footnote{Observe that the partition has all parts
$\geq 2$ since $m_{1/2}=2$}
       
\bigskip

Now consider a partition(according to our definition) of the integer number $A$ which represents 
the area of our isolated black hole. Since we are considering our minimum allowed spin $j$ to be
$1/2$, such that $m_{1/2}=2$, we want to consider partitions of the number $A$ such that its parts are in terms of natural numbers greater or equal to $2$.

Then $A$ is expressed 
$A= \lambda_1 + \lambda_2 +...+\lambda_k$, where
$\lambda_1 \geq \lambda_2 \geq.....\geq \lambda_k \geq 2$

\bigskip

Then we just think of the integers $\lambda_{j}$ as $m_{j}$. Since some of the $\lambda_{j}=m_{j}$
may be repeated it is clear that we will have a decomposition of the number $A$ as in formula
$(6)$.
 
\bigskip

This implies then that the number of indistinguishable configurations $\mathcal{N}$ which account for a fixed area $A$
equals the number of partitions
of $A$ with all parts $\geq 2$.  An exercise in number theory courses
shows that this number is given by
$p(A)-p(A-1)$.

\bigskip

It is easy to notice that for the case of equidistant spectrum given by $A=  2 \sum_{i=1}^{n} j_{i}$ 
where $j$ is a half integer $j=m/2$ the number of configurations which account a fixed area $A$ 

\begin{equation}
A=  n_{1/2} m_{1/2} +  n_{1} m_{1}
+...+ n_{s_{max}}m_{s_{max}/2} 
\end{equation}
where the minimum $m_{1/2}=1$ is given by the number of partitions of $A$,
that is $p(A)$.

We therefore have the following. If we denote by $\mathcal{N}_{A_{(j+1/2)}}$ the number
of indistinguishable configurations which account a fixed area for the case of equidistant spectrum
$(j+1/2)$ and  $\mathcal{N}_{A_{j}}$ the number
of indistinguishable configurations which account a fixed area for the case of equidistant spectrum
$j$, we have that

\begin{equation}
\mathcal{N}_{A_{j}} > \mathcal{N}_{A_{(j+1/2)}}
\end{equation}

\bigskip

Now for the case of spectrum $A_{j}= 2 \sqrt{j (j+1)}$ it is known that counting configurations
is not restricted to an exact sum but the sum of any configuration gives a number between
an interval $[A-\delta, A+\delta ]$. As the spin $j=m/2$ for $m$ integer $\geq 1$ we can write
$A_{j}=A_{m}=\sqrt{m (m+2)}=\sqrt{m^{2}+2m}$. With this kind of spectrum we can only expect that any configuration will give an irrational number area. 
One way to go around the problem
would be to consider $A_{j} \sim (m+1)$ where counting the number of 
indistinguishable configurations
will lead us to the same result we obtained when considering the spectrum $A_{j}=(j+1/2)$.

\bigskip

In Loop Quantum Gravity the number of different indistinguishable configurations $\mathcal{N}$ 
which account for a fixed area $A$ of an isolated horizon is(at least when $j$ is large)
given 
by $\mathcal{N}=p(A)-p(A-1)$.

We could say that the asymptotic behaviour is given by

\begin{equation}
\mathcal{N} \sim \frac{1}{4\sqrt{3}} \bigg[ \frac{1}{A} \exp (\pi \sqrt{2A/3}) 
- \frac{1}{(A-1)} \exp (\pi \sqrt{2(A-1)/3})\bigg]
\end{equation}
where in this last formula we have used the 
asymptotic behaviour of $p(N)$ known in number theory.

We should say that when counting different indistinguishable configurations, 
the entropy is asymptoticly dominated by the square root of the area. In \cite{kk1} 
a counting calculation of indistinguishable configurations shows a 
similar behaviour to ours. 

In \cite{bcc} a similar formula is given. The difference and contribution in our paper
is that we are dealing with the equidistant spectrum
and we are also pointing to the number theory methods which we believe are so related to
the counting of states of black hole entropy in loop quantum gravity. 

For instance in \cite{abpbv} a deep relation to number theory is given. In a future work we 
plan to deal with the case of the real irrational spectrum of area of loop quantum gravity
pointing to more relations to number theory methods. 

For instance it is easy to observe that if  
$\mathcal{N}_{A_{(j+1/2)}}$ denotes the number
of indistinguishable configurations which account a fixed area for the case we treated here of equidistant spectrum
$(j+1/2)$ and  $\mathcal{N}_{A_{\sqrt{(j(j+1))}}}$ the number
of indistinguishable configurations which account a fixed area for the case of 
the original spectrum of loop quantum gravity then

\begin{equation}
\mathcal{N}_{A_{\sqrt{(j(j+1))}}} < \mathcal{N}_{A_{(j+1/2)}}
\end{equation}
which shows that the theory of partitions is essential for the counting.

We are therefore showing that these calculations are profoundly
related to number theory. We propose to study in a future work a mathematical rigorous treatment
on these relations between counting black hole states in loop quantum gravity 
and analytic number theory.

\bigskip

\danger{Acknowledgement} This work was supported by CONACYT proyect called
"PROBLEMAS MATEMATICOS DE LA FISICA CUANTICA"

\end{document}